\documentclass[superscriptaddress,reprint,amsmath,amssymb,aps,prl
]{revtex4-2}

\usepackage{rotating}

\usepackage{natbib}
\usepackage{amsmath,amssymb}% Include figure files
\usepackage{graphicx}% Include figure files
\usepackage{dcolumn}% Align table columns on decimal point
\usepackage{bm}% bold math

\newcommand{\beq}{\begin{equation}}
\newcommand{\eeq}{\end{equation}}

\begin{document}

\title{Stochastic Heating in the Sub-Alfv\'{e}nic Solar Wind}

\author{Trevor A. Bowen}\email{tbowen@berkeley.edu}
\affiliation{Space Sciences Laboratory, University of California, Berkeley, CA 94720-7450, USA}
\author{Tamar Ervin }
\affiliation{Space Sciences Laboratory, University of California, Berkeley, CA 94720-7450, USA}
\affiliation{Physics Department, University of California, Berkeley, CA 94720-7300, USA}
\author{Alfred Mallet}
\affiliation{Space Sciences Laboratory, University of California, Berkeley, CA 94720-7450, USA}
\author{Benjamin D.~G. {Chandran}}
\affiliation{Department of Physics \& Astronomy, University of New Hampshire, Durham, NH 03824, USA}
\author{Nikos Sioulas}
\affiliation{Space Sciences Laboratory, University of California, Berkeley, CA 94720-7450, USA}
\author{Philip A. {Isenberg}}
\affiliation{Department of Physics \& Astronomy, University of New Hampshire, Durham, NH 03824, USA}

\author{Stuart D. Bale}

\affiliation{Space Sciences Laboratory, University of California, Berkeley, CA 94720-7450, USA}
\affiliation{Physics Department, University of California, Berkeley, CA 94720-7300, USA}

\author{Jonathan Squire}
\affiliation{Department of Physics, University of Otago, 730 Cumberland St., Dunedin 9016, New Zealand}

\author{Kristopher G. Klein}
\affiliation{Lunar and Planetary Laboratory,University of Arizona, Tucson, AZ}

\author{Oreste Pezzi}
\affiliation{Institute for Plasma Science and Technology, National Research
Council of Italy, G. Amendola 122/D, Bari, I-70126, Italy}%https://orcid.org/0000-0002-7638-1706
\begin{abstract}
Collisionless dissipation of turbulence is important for heating plasmas in astrophysical, space physics, and laboratory environments, controlling energy, momentum and particle transport. We analyze Parker Solar Probe observations to understand the collisionless heating of the sub-Alfv\'{e}nic solar wind, which is connected to the solar corona. Our results show that linear resonant heating through parallel-propagating cyclotron waves cannot account for turbulent dissipation in sub-Alfv\'{e}nic region,  which observations suggest may dissipate turbulence at distances further from the Sun. Instead, we find that stochastic heating can account for the observed ion energization; however, because the dominant contributions arise from infrequent, large‐amplitude events, turbulent intermittency must be explicitly incorporated. These observations directly connect stochastic heating via breaking of the proton magnetic moment with the intermittent and inhomogeneous heating of turbulence reported in many previous studies. Our identification of stochastic heating as a dynamic mechanism responsible for intermittent heating of the solar wind has significant implications for turbulent dissipation in the lower corona, other astrophysical environments, and laboratory plasmas.

\end{abstract}

\maketitle

The collisionless heating of the solar wind and corona is a well known phenomenon \cite{Marsch1982a, CranmervanBallegooijen2003} with important implications for dissipation and heating of other astrophysical environments \cite{Quataert1998,Kunz2011,Zhuravleva2014,Ressler2015,Howes2024} as well as laboratory and fusion plasmas  \cite{Abel2013,Helander2015,Howes2018}. Remote observations of the inner corona reveal significant temperature anisotropy with $T_\perp>T_\parallel$ \cite{Kohl1997,Kohl1998,Cranmer1999,Cranmer2000}, a common feature of collisionless plasmas.
Residual anisotropic heating is additionally present in the solar wind \cite{Marsch1982b,Hellinger2011}, resulting in non-adiabatic temperature profiles with distance \cite{Richardson1995,Hellinger2013,Zaslavsky2023,Bowen2025b}. These observations have led to significant interest in mechanisms that preferentially heat plasmas perpendicular to the magnetic field direction, which can be broadly separated into either quasilinear resonant electromagnetic field-particle interactions \cite{Kennel1966,Marsch1982a,Leamon1998b,Leamon1999} or non-resonant mechanisms \cite{JohnsonCheng2001,Chen2001,Chandran2010,Mallet2017,Loureiro2017,Vech2018} . 

 Recent observations from Parker Solar Probe (PSP) \cite{Vech2021,Bowen2022,Bowen2024b,Shankarappa2024} have highlighted the importance of quasilinear (QL) heating through left-hand polarized electromagnetic fields at ion-kinetic scales, which are likely quasi-parallel ion-cyclotron waves ($\parallel$-ICWs)\cite{Jian2010,Boardsen2015,Wicks2016,Bowen2020a,Bowen2020d,Liu2023}.  Similar observations provide evidence for cyclotron heating in the 1 AU solar wind \cite{He2015,Woodham2018,Zhao2020,Telloni2019} and turbulent magnetosheath \cite{Afshari2024}, suggesting that quasilinear resonant ICW interactions play an important role in collisionless dissipation in a range of environments. While there is a growing consensus regarding a role of $\parallel$-ICWs in mediating turbulent heating \cite{Bowen2024a}, there is a long standing debate regarding whether the observed $\parallel$-ICWs are directly responsible for the dissipation of turbulence or if they are byproduct of other mechanisms \cite{Chandran2010b,Squire2022}. This important debate largely stems from the tendency of Alfv\'enic turbulence to transfer energy efficiently to high perpendicular wave-number, $k_\perp$, rather than to high parallel-wavenumber, $k_\parallel$ \cite{Shebalin1983,GS95,Bieber1996,Horbury2008,Chen2012,Mallet2015,Duan2021}, leading to anisotropic fluctuations at small scales with $k_\perp\gg k_\parallel$; this type of cascade cannot directly produce $\parallel$-ICW. It is often thought that $\parallel$-ICWs are generated via the $T_\perp>T_\parallel$ Alfv\'{e}n/ICW instability \cite{Gary1993}, which cools the plasma \cite{Hellinger2006}. Observations \cite{Marsch2001,Heuer2007,He2015,Bowen2022} show that the ion velocity distribution function (VDF) is roughly constant along the $\parallel$-ICW diffusion contours \cite{IsenbergLee1996}, implying marginal stability and further complicating observation and interpretation of the dynamics. %Identifying the correct interpretation is often difficult due to the marginal stability of the plasma \cite{Marsch2001,Heuer2007,He2015,Bowen2022} along the ICW diffusion contours \cite{IsenbergLee1996}. 

%Furthermore, the tendency of turbulence to form perpendicular structure as the cascade transfers energy to consecutively smaller spatial scales \cite{Shebalin1983,Bieber1996,Horbury2008,Chen2012,Duan2021} may preclude linear cyclotron resonant interactions and the formation of parallel propagating ICWs via turbulence. 

If resonant interactions cannot supply the requisite levels of heating for turbulent dissipation, other, inherently nonlinear, processes must take an active role. Such mechanisms, often termed ``stochastic'' occur in a variety of contexts: cosmic ray transport \cite{Kempski2023}, diffusive shock acceleration \cite{Drury1983}, turbulent dissipation in the magnetopause \cite{JohnsonCheng2001}, drift-wave turbulence \cite{McChesney1987,McChesney1991}, and energetic ions in tokomak experiments \cite{Hauff2009}. In the solar wind and corona, stochastic heating (SH) of ions, understood as a random walk in energy spurred by repeated non-resonant interactions with turbulent fluctuations \cite{Chandran2010,Hoppock2018,Isenberg2019}, has been suggested as viable plasma heating mechanism \citep{Bourouaine2013,Xia2013,Klein2016,Vech2017,Martinovic2019,Martinovic2020}. %The SH mechanism is related to conservation of the 

%large-amplitude fluctuations in the electric field cause ion orbits to become chaotic, breaking the conservation of the magnetic moment $\mu=m v_\perp^2/B$,  has been suggested as viable means for heating in the sub-Alfv\'{e}nic solar wind and corona \citep{Bourouaine2013,Xia2013,Klein2016,Vech2017,Martinovic2019,Martinovic2020}. %This mechanism, which
%As such, nonlinear processes may take an important role in particle transport and heating. Stochastic mechanisms that result in turbulent dissipation and particle transport appear

 In contrast to the canonical formalism of QL theory \cite{Kennel1966}, stochastic heating is highly sensitive to intermittency in the turbulence \cite{Mallet2019,Cerri2021}, which is understood as the the broad, heavy-tailed, distribution of fluctuation amplitudes that intrinsically formed by turbulence \cite{Burlaga1991,Horbury1997, Sorriso-Valvo1999}. Importantly, correlations between intermittency and heating are widely recognized \cite{Greco2012,Osman2011,Osman2012,Chasapis2015,Matthaeus2015,Qudsi2020}, which suggests that QL resonant mechanisms that are canonically derived under non-intermittent conditions (e.g. \cite{Kennel1966}), cannot entirely explain turbulent heating. Constraining the interplay of QL mechanisms, which are important in understanding the observed wave-populations \cite{Bowen2022,Verniero2022}, with intermittent non-resonant mechanisms, such as SH, is imperative in understanding dissipation \cite{Mallet2019}. 
 
 Here, we use PSP  \cite{Fox2016} observations to investigate heating of the sub-Alfv\'{e}nic solar wind, where the Alfv\'{e}n speed exceeds the solar wind speed \cite{Kasper2021}. In contrast to previous results from larger heliocentric distances \cite{Bowen2022,Bowen2024b}, our results show the plasma is locally unstable to $\parallel$-ICWs \cite{Gary1993,Gary2004}, but with a net energy transfer that is far below the local turbulent cascade rate. Thus $\parallel$-ICW interactions are not energetically relevant in bulk heating of this sub-Alfv\'{e}nic solar wind stream. While the $\parallel$-ICWs cannot explain turbulent dissipation, we find SH rates \cite{Chandran2010,Klein2016} that are consistent with the turbulent cascade rate \cite{MarschTu1997,Sorriso-Valvo2018}, if (and only if!) we account for the intermittent nature of the turbulence. 
These observations suggest a connection between SH via magnetic moment breaking \cite{Chandran2010} and many studies that highlight intermittent and inhomogeneous turbulent heating \cite{Greco2012,Osman2011,Osman2012,Chasapis2015,Matthaeus2015,Qudsi2020}. Conversely, the fact that intermittency is required to attain SH rates that match the turbulent cascade implies that non-intermittent fluctuations may be irrelevant in heating. These results provide strong constraints on collisionless dissipation in Alfv\'{e}nic turbulence with implications for coronal heating \cite{Cranmer2007,Chandran2010b,Ballegooijen2017}.

%we find signatures of SH via magnetic moment breaking by turbulence \cite{Chandran2010,Hoppock2018}, %We compare SH rates to the observed QL heating rates of parallel-propagating ion cyclotron waves \cite{Kennel1966,Isenberg2007}.  

\begin{figure}
    \centering
    \includegraphics[width=\linewidth]{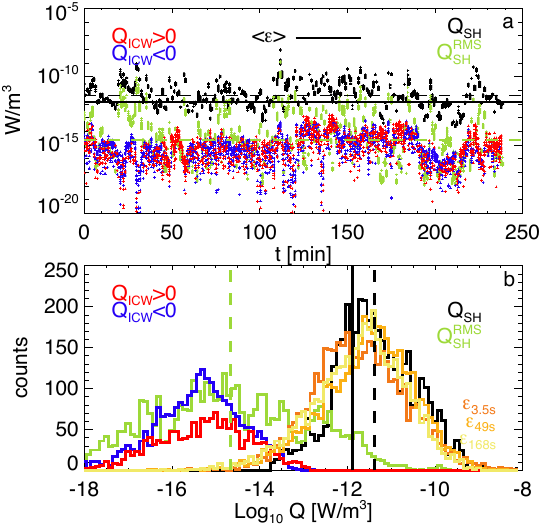}
    \caption{a) Heating rates as function of time: stochastic heating $Q_{SH}$ black, non intermittent stochastic heating $Q^{RMS}_{SH}$ (green), ICW emission $Q_{ICW} <0$ (blue), ICW heating $Q_{ICW} >0$ (red). Median values for $Q_{SH}$  and  $Q^{RMS}_{SH}$  are shown in dashed lines. Solid black line shows the median local-energy transfer (LET) estimate $\langle\epsilon\rangle$ of turbulent cascade rate. b) Distributions of heating rates of data shown in top panel; additionally distributions of $\epsilon$ from LET estimate at lags corresponding to inertial range scales with lags at 3.5s, 49s, and 168s are shown (orange, gold, yellow). }
    \label{fig:1}
\end{figure}
\paragraph{SH Model.---}   Studies of SH in Alfv\'{e}nic turbulence show strong sensitivity to the fluctuation amplitude \cite{White2002,Voitenko2004,Bourouaine2008,Cerri2021}.  The efficiency of SH is quantified via a stochasticity parameter, $\epsilon_u=\delta u/v_{\perp}$ \cite{Chandran2010}, where $\delta u$ is the turbulent fluctuation amplitude at a particle's gyroradius, $\rho_i$, the perpendicular speed $v_{\perp}=\rho_i\Omega_i$ with $\Omega_i=qB/m_i$. %The SH of the solar wind and corona has been studied largely in the context of Alfv\'{e}nic turbulence \cite{Chen2001,White2002,JohnsonCheng2001,Bourouaine2008,Chandran2010}, which has highly correlated turbulent magnetic and velocity field fluctuations and is regularly observed by PSP\cite{McManus2020,Chen2020,Shi2021}. 
%Where MHD breaks down, evidence suggests that the MHD-inertial range turbulence couples into a kinetic Alfv\'{e}ni-wave (KAW) cascade \cite{Bale2005,Chen2010a,Salem2012}. %When turbulent fluctuations become similar in size to the ion-gyroscale, $\rho_i=\frac{mv_{ith\perp}}{qB}$, they can perturb a particle's magnetic moment, which is generally conserved in the limit of slowly varying fields in the presence of small fluctuation amplitudes. 
%The simplified heating rate at the thermal gyroradius, $\rho_{th}$ from stochastic heating as developed in \cite{Chandran2010} \begin{equation}
%Q^{\rho_{th}}_{{SH}} =c_1\frac{\delta v_{\rho_{th}}^3}{\rho_{th}}e^{-c_2\frac{v_\perp}{\delta v_{\rho_{th}}}}\end{equation} which omits dependence on velocity space.
 Chandran et al.~2010 (hereafter, C+2010, \cite{Chandran2010}) estimate the SH energy diffusion coefficient of low-$\beta$ Alfv\'{e}nic turbulence as \begin{align}D_E=c_1v_{\perp}\delta u^3\Omega_ie^{\frac{-c_2}{\epsilon_u}} \label{eq:DE}\end{align}
with constants $c_1$, $c_2$ of order unity and a corresponding velocity-space diffusion coefficient $D_\perp=D_E/v_{\perp}^2$. The exponential in Eq.~(\ref{eq:DE}) results in extreme sensitivity to $\epsilon_u$ such that SH is irrelevant for $\delta u\ll v_\perp$\cite{Cerri2021}. While $D_E$ is a function of $v_{\perp}$, C+2010 specifically focus on SH at the thermal gyroradius $\rho_{th}=v_{th\perp}\Omega_i$ noted as $Q^{\rho_{th}}_{{SH}}$, with
\begin{equation}\label{eq:Qthsh}
Q^{\rho_{th}}_{{SH}} =c_1\frac{\delta u_{\rho_{th}}^3}{\rho_{th}}e^{-c_2/\epsilon_{\rho_{th}}},\end{equation}
where $v_{th\perp}$ is the perpendicular ion-thermal speed, $\delta u_{\rho_{th}}$ is the fluctuation amplitude at the thermal gyroscale, and $\epsilon_{\rho_{th}}= \delta u_{\rho_{th}}/v_{th\perp}$.
Observational studies have used values of $c_1\approx 0.75$ and $c_2\approx0.34$ \cite{Martinovic2019,Martinovic2020}, based on test particle simulations \cite{Chandran2010}. Alternatively, matching the SH rate, $Q_{SH}$, to the turbulent cascade rate, $\epsilon^T$, \cite{Bourouaine2013} or via numerical simulations \cite{Xia2013} can constrain the coefficients.

While the C+2010 expression for SH in Eq.~(\ref{eq:Qthsh}) only considers heating localized at the thermal-perpendicular speed, more general SH formalisms \cite{Klein2016,Cerri2021} express SH as a $v_\perp$-dependent diffusion of a gyrotropic VDF, %$f(v_\perp,v_\parallel)$ and associated 
with%as
\begin{align}
    %\frac{\partial f}{\partial{t}}&=\frac{1}{v_\perp}{\frac{\partial}{\partial v_\perp}}\left(v_\perp D_\perp \frac{\partial f}{\partial v_\perp}\right)\\
Q_{SH}&=\frac{m_i}{2}\int v_\perp^2 \frac{\partial f}{\partial{t}} d^3v= -2 \pi m_i\int_0^\infty v_\perp^2 D_\perp \frac{ \partial F(v_\perp)}{ \partial v_\perp}  dv_\perp,\label{eq:QSH} \end{align}
where the reduced VDF is %$d^3v=2\pi v_\perp dv_\perp dv_\parallel$ only considering perpendicular heating such that 
$F(v_\perp)= \int f(v_\perp,v_\parallel) d v_\parallel$ and the last term in Eq.~(\ref{eq:QSH}) is obtained by integration by parts. We also define $Q_{SH}=\int dQ_{SH}$.
%$$Q_\perp=\int (m_i/2)v_\perp^2  1/v_\perp \frac{ \partial}{ \partial v_\perp} \left(v_\perp D_\perp \frac{ \partial}{ \partial v_\perp}F(v_\perp)\right) 2\pi v_\perp dv_\perp$$
%$$Q_{SH}=\int \pi m_iv_\perp^2 \frac{ \partial}{ \partial v_\perp} \left(v_\perp D_\perp \frac{ \partial}{ \partial v_\perp}F(v_\perp)\right)  dv_\perp$$
%Integration by parts gives 
%\begin{equation}\label{eq:QSH} Q_{SH}= -2 \pi m_i\int_0^\infty v_\perp^2 D_\perp \frac{ \partial F(v_\perp)}{ \partial v_\perp}  dv_\perp.\end{equation}
An ion with perpendicular velocity $v_\perp$ has a gyroradius $\rho = v_\perp/\Omega$, and thus $F(v_\perp)$ can equivalently be considered a distribution of gyroradii corresponding to perpendicular speeds via $\rho_i = v_\perp / \Omega_i$. The quantity $\delta u$ in Eq.~(1) corresponds to the turbulent amplitude at the gyroradius of the particles with $v_\perp$, and thus can also be considered a function of $v_\perp$. While scale-dependent SH has been applied in both analytic theory and simulation \cite{Klein2016,Cerri2021}, we presently apply the scale dependent approach to in situ observations. Importantly, extending $D_E$ strictly based on C+2010 is valid if $v_\perp>v_{th\perp}$, as these particles are heated by fluctuations at scales $\rho>\rho_{th}$, and the MHD electric field is appropriate \cite{Klein2016}. It is technically not valid to extend it to lower $\rho<\rho_{th}$ since the scaling of the electric field fluctuations changes for scales below the thermal ion gyroradius (e.g. the Hall field); however, as we will show, minimal heating occurs when $\rho<\rho_{th}$.

\paragraph{Methods}We calculate a scale dependent fluctuation amplitude of the magnetic field $\delta{B}$ using 5-point increments \cite{Cho2019,Sioulas2024} over a range of 88 time lags ranging from 0.85 ms $<\tau< 7.23 $ s. The corresponding frequencies range from $0.07 <f <586$ Hz.
%The wavelet transform is used to construct a power-spectral density $S_B(f)$ (see end matter).
The lag $\tau$ is related to spacecraft frequencies as $f=1/2\tau$ \cite{Cho2009}; the corresponding physical spatial scales are determined using the modified Taylor hypothesis \cite{Klein2015}, which is appropriate for large amplitude-outward propagating Alfv\'{e}nic fluctuations, and necessary for the sub-Alfv\'{e}nic region where $V_A>V_{sw}$ \cite{Bourouaine2019,Chhiber2019}:
\begin{align} \label{eq:modtay}
2\pi f= k_\parallel V_A +k_\parallel V_{sw} cos\theta_{VB}+k_\perp V_{sw} sin\theta_{VB}.
%2\pi f_{str}= k_\perp V_{sw} sin\theta_{VB}\\
\end{align}

%This expression allows us to apply necessary correction to convert from magnetic field to velocity fluctuations (as diffusion coefficients rely on $\delta u$ rather than $\delta B$).
The Doppler shift equation can be used to connect perpendicular fluctuations with $k_\perp\rho=1$ to $f$ as $2\pi f= k_\perp V_{sw} sin\theta_{VB}$, which enables conversion from $\delta B$  to $\delta u$ assuming Alfv\'{e}nic turbulence and a transition to the KAW mode at $\rho_{th}$ such that $\delta u=\delta B/B_0 v^{kaw}_{ph}$
where $v^{kaw}_{ph}=v_A\sqrt{1+k_\perp^2 \rho_{th}^2}$ \cite{Hollweg1999}. This correction is valid for frequencies $\omega <\Omega_i$, corresponding to the largest sub-ion kinetic scales, and for highly imbalanced turbulence, such as this stream \cite{Zank2022,Zhao2022}.

%$S_{u}(f,t)={S_B}(1+k_\perp^2 \rho_{th}^2)$ \cite{Hollweg1999}. The amplitude is computed as the integral
%$\delta u(k)_=\int_k^{\infty} P_u dk$. %We use the $\delta u_\rho$ to indicate the amplitude 
%as a scale 
%$\delta u_\rho=\delta B_\rho/B_0 v^{kaw}_{ph}$
%$where $v^{kaw}_{ph}=v_A\sqrt{1+k^2 \rho_{th}^2}$ \cite{Hollweg1999}. 

%using 5-point structure functions \cite{Cho2019,Sioulas2024} over a range of 88 time lags ranging from 0.85 ms $<\tau< 7.23 $ s, corresponding to a range of frequencies from $0.07 <f <585$ Hz. %The instantaneous turbulent amplitude is then given as $|{\delta}B(t,\tau)|$. These lags and the wavelet spectra given by $\tilde{B}=2\Delta t\langle|{\delta}B(f,t)|^2 \rangle$ where the brackets denote the time average and $\Delta t$ is the sampling rate \cite{DudokdeWit2013,Bowen2024b}.

While we neglect the $k_\parallel$ term for the KAW conversion (as anisotropy $k_\perp\gg k_\parallel$ is particularly strong at kinetic scales \cite{Duan2021}), contributions from $k_\parallel$ are significant in converting observations at lag $\tau$ to spatial scales in the sub-Alfv\'{e}nic wind (\cite{Klein2015}, see end matter ). Assuming a spectrum of critically balanced turbulence \cite{GS95,Mallet2015}, $k_\perp\delta u=k_\parallel V_A$, uniquely relates fluctuations at lag $\tau$ to a specific $k_\perp$ {(see end matter)}. Then, requiring $k_\perp \rho=1$ associates each $\tau$ with a single $v_\perp=\Omega_i/k_\perp$ (computed for each interval). We can then construct $\epsilon_u (f,t)$ from the increment estimate of the amplitude and $v_\perp$ allowing for an estimate of $D_E$ in Eq. (\ref{eq:DE}). We note that the importance of the  $k_\perp\rho=1$ is an extension of the original C+2010 formalism, whereas recent extensions of SH suggest that time scales (rather than spatial scales) may be most important \cite{Johnston2024,Mallet2025b}. %Further discussion, and caveats, are included in the end matter.

%$\lambda$ is \beginD_E (t,\lambda)=v_{\perp\lambda} \delta u^3_\lambda e^{-c
%The wavelet amplitude is then gaiven as $|{\delta}B(\lambda,t)|$  and the $\delta u_\lambda$ fluctuation is constructed assuming that the turbulent fluctuations are Alfv\'{e}nic and transition to the KAW mode at the thermal gyroscale $\rho_{th}$ such that $\delta u_\lambda=\delta B_\lambda/B_0 v^{kaw}_{ph}$
%where $v^{kaw}_{ph}=v_A\sqrt{1+k^2 \rho_{th}^2}$.
Turbulence is inherently intermittent \cite{Matthaeus2015}, which is important to the effectiveness of SH \citep{Mallet2019} through locally enhancing $\epsilon_u$ \cite{Cerri2021}, which can counter the exponential suppression in Eq.~(\ref{eq:DE}). The importance of intermittency also arises in the third-order dependence of $D_E\propto \delta u^3$, which indicates an inherent sensitivity to large amplitude, rare fluctuations.
In calculating SH rates, we consider two different estimates of $D_E$. The usual estimate, $D^{rms}_E(f)$, is constructed from the rms $\delta u^{rms}=\sqrt{\langle \delta u^2\rangle}$ amplitude over a time interval, which averages out intermittent effects. In the second, $D_E$ is calculated individually for each increment $ \delta u(t,f)$, giving an ensemble $D_{E}(t,f)$, which includes the effects of intermittency. The time-average $\bar{D}_E=\langle |D_E|\rangle$ then includes intermittency in $\delta u$. This method amounts to treating $Q_{SH}$ and $D_E$ as functions of the random variable $\delta u$ with average quantities integrated over the distribution $\langle X\rangle= \int X(\delta u) P(\delta u)d\delta u$.%; this approach is essentially a velocity-space dependent of the theoretical work in \cite{Mallet2019}.
%\paragraph{Distribution Functions} Estimate of the heating rate $Q_{SH}$ requires accurate measurement of gradients in the distribution function, which we obtain via the PSP/SPAN-i instrument \cite{Kasper2016,Livi2022}. We apply drifting-biMaxwellian fits following techniques adapted for PSP \cite{Verniero2020,Klein2021,Bowen2022}. %We also investigated a nonparametric radial-basis function interpolation \cite{broomhead1988,Bowen2024b} but found no variation with choice of representation and thus proceed with the more commonly used drifting beam-core biMaxwellian fit VDFs.

\paragraph{Data \& Results}

We study a stream from April 8th 2021 spanning 10:00:00-14:00:00, which was identified as the first prolonged encounter of PSP with sub-Alfv\'{e}nic solar wind \cite{Kasper2021}. Several studies have investigated the turbulent properties of this interval \cite{Zank2022,Zhao2022,Bandyopadhyay2022}, which has $\beta \approx 0.05$ and is highly imbalanced\cite{Zhao2022}, suggesting applicability of the C+2010 SH mechanism. %Estimate of the heating rate $Q_{SH}$ requires accurate measurement of gradients in the distribution function, which we obtain via the PSP/SPAN-i instrument a\cite{Kasper2016,Livi2022}. We apply drifting-biMaxwellian fits following techniques adapted for PSP \cite{Verniero2020,Klein2021,Bowen2022}. %While the E8 sub-Alfénic stream has been interpreted as crossing into the solar corona\cite{Kasper2021}, the ``Alfv\'{e}ni'' region, where the wind speed approximates that of the Alfv\'{e}nic speed, may be fragmented and irregular \cite{Chhiber2022} or affected by coronal mass ejections \cite{Romeo2023}, such that it is unclear whether PSP is properly in the corona.

\begin{figure}
    \centering
    \includegraphics[width=\linewidth]{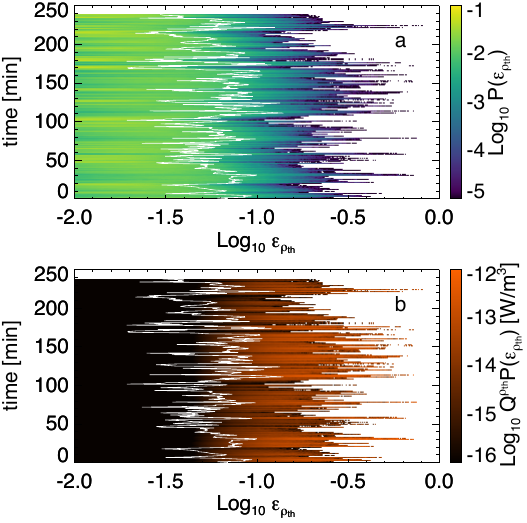}
    \caption {a) Spectrogram showing distribution of normalized amplitude $P(\epsilon_{\rho_{th}})$ evaluated at the thermal proton gyroscale $\epsilon_{\rho_{th}}=\delta u_{\rho_{th}}/v_{th\perp}$ in each interval. b) Distribution of normalized amplitudes weighted by SH rate $P(\epsilon_{\rho_{th}})Q^{\rho_{th}}_{SH}$, showing that heating occurs in the tails. The white contour shows the largest 25\% level of $P(\epsilon_{\rho_{th}})$.}
    \label{fig:2}
\end{figure}
 
The interval is broken into non-overlapping sub-intervals lasting 55.92 seconds ($\approx 1500\Omega_i$), which is sufficiently long enough to calculate SH rates well into the inertial range. Increment analysis is applied to each interval to construct $\bar{D}_E$ and $D^{rms}_E$. In each sub-interval, the $\parallel$-ICW spectrum is computed by filtering out wavelet coefficients without strong left-hand polarization. The $\parallel$-ICW heating rate $Q_{\parallel ICW}$ can be empirically calculated from the $\parallel-$ICW spectrum and the gradients of the VDF \cite{Kennel1966,Bowen2022,Bowen2024b}, which we obtain via the PSP/SPAN-i instrument \cite{Kasper2016,Livi2022} at a 3.5s cadence. %Proton VDFs are measured by SPAN-i every 3.5 seconds. %For each SPAN-i VDF in the 55.92 second interval. 
We apply drifting-biMaxwellian fits \cite{Verniero2020,Klein2021,Bowen2022} to each VDF in each sub-interval.
%A total of 4111 SPAN-i VDFs were studied.
The heating rates $Q_{SH}$, $Q_{SH}^{RMS}$, and $Q_{\parallel ICW}$ are then computed (see end matter for further discussion).

Figure 1(a) shows the measured $Q_{SH}$ and $Q_{SH}^{RMS}$ for the drifting biMaxwellian fits to the VDF from  $\bar{D}_E$ and $D^{RMS}_E$. Median estimate of heating are shown as dashed lines, the median $Q_{SH}$ rate is similar to the median turbulent cascade rate $ \epsilon^T$ computed via local energy transfer (LET, \cite{Sorriso-Valvo2018}, see end matter). The absolute value of $Q_{ICW}$ is plotted in red when $Q_{ICW}>0$ (heating) and blue when $Q_{ICW}<0$ (emission). Throughout, SH dominates over $\parallel$-ICW heating, with $|Q_{ICW}|\ll Q_{SH}$. %indicating that the SH mechanism is dominating over the parallel ICW interactions. 
Overall, there is net $\parallel$-ICW emission in the interval suggesting that waves are generated by the A/IC instability. Oscillating positive and negative $Q_{ICW}$ suggests the plasma is in a marginally stabile state with respect to the A/ICW instability.%., which is in contrast to several previous studies \cite{Bowen2022,Bowen2024b,Shankarappa2024}) that found net ICW resonant heating on order the turbulent cascade rate. The oscillation between positive and negative vales of $Q_{ICW}$ suggests the plasma is close to marginal stability state.

 Fig 1b) shows probability distributions of the various heating rates. The SH rates computed from $D_E^{rms}$ are more than an order-of-magnitude lower than those computed from $\bar{D}_E$, which includes intermittency. Our means of including intermittency via averaging over $D_E$ differs from that previously applied in numerical work of \cite{Cerri2021}, but reveals the significant enhancement to the SH rate due to intermittent fluctuations \cite{Mallet2019}.  %These results show that when intermittency is included, SH is capable of providing the necessary heating to the solar wind, this is largely consistent with previous observations of intermittent heating\cite{Osman2011,Osman2012,Qudsi2020}.
  
To further highlight the role of intermittency in SH, Figure 2(a) shows the probability distribution of the stochasticity parameter measured at the thermal gyroscale, $P(\epsilon_{\rho_{th}})$, in each interval (the single scale is chosen for illustrative purposes). Figure 2(b) shows $P(\epsilon_{\rho_{th}})$ weighted by the heating rate at the thermal gyroradius, Eq.~(\ref{eq:Qthsh}), i.e.  $P(\epsilon_{\rho_{th}})Q_{SH}^{\rho_{th}}$, which demonstrates the level of heating associated with a given $\epsilon_{\rho_{th}}$. The heating very clearly occurs in the intermittent, large-amplitude tails of the distribution\cite{Matthaeus2015}, which have occurrence probabilities $\lessapprox 0.1\%$. The upper quartile is shown in white.

 \begin{figure}
    \centering
    \includegraphics[width=\linewidth]{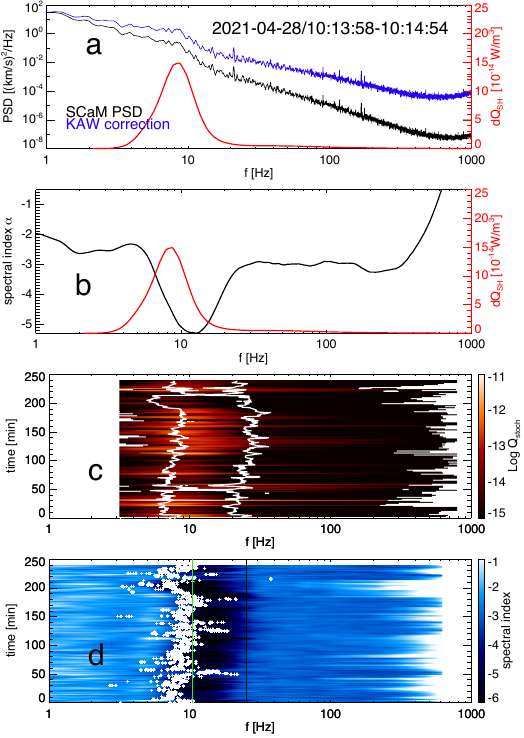}
    \caption{a) PSD constructed from PSP/FIELDS magnetic field from 2021-04-28/10:13:58-10:14:54 \cite{Bale2016,Bowen2020b}; PSD is normalized using Alfv\'{e}nic speed and the KAW normalization has been applied in blue; differential heating $dQ_{SH}(f)$ is shown in red. b) Moving window spectral index $\alpha$ computed from (non KAW normalized) PSD. c) Spectrogram showing  $dQ_{SH}$ as function of frequency and time a white contour shows where spectral index of fluctuations  has $\alpha <-4$. d) Moving window spectral index $\alpha$ as function of frequency and time; white dots show the $f$ with the maximum $dQ_{SH}.$ Black line shows average  $f=V_{sw}/2\pi\rho_i$; green line shows average $f=V_{sw}\text{sin}\theta_{VB}/2\pi\rho_i$}
    \label{fig:3}
\end{figure}
% Only when including intermittent effects is SH capable of providing the necessary heating to the solar wind, . Distributions of LET rates $\epsilon$ are also shown and coincide with the distributions of $Q_{SH}$ as computed with intermittent effects ($\bar{D}_E$). %r Figure 2(c) additionally shows PDF of the heating rate computed by RBFs vs drifting biMaxwellians. The drifting biMaxwellian approximation tends to have  heating rate of about 30-70\% of the RBF approximation. 

 Figure \ref{fig:3}(a) shows the power spectral density from the SCaM data converted to Alfv\'{e}nic units for a single sub-interval, the KAW correction to the spectra is additionally illustrated. We also plot the scale dependent $dQ_{SH}$. Heating is observed to peak in the transition range \cite{Sahraoui2009,Kiyani2009,Sahraoui2010}, which has been interpreted as a signature of dissipation \cite{Denskat1983,Smith2006,Bowen2020c}. Figure \ref{fig:3}(b) shows an estimate of the moving spectral index of the turbulent spectra, $\alpha$: again, it is clear that SH becomes effective at frequencies associated with spectral steepening. Figure \ref{fig:3}(c) shows a spectrogram of $dQ_{SH}$. A white contour is shown outlining the range of scales for which the spectral index is very steep with $\alpha<-4$. Over the entire interval, it is clear that the SH occurs at the range of scales corresponding to transition range steepening. Figure \ref{fig:3}(d) shows the estimate of the spectral index, white dots mark the location of the peak SH value, which occurs in the low-end of the transition range.  Figure \ref{fig:3}(d) also shows the  frequency corresponding to the average gyroscale $f=V_{sw}/2\pi\rho_i$ as well as the average $f=V_{sw}\text{sin}\theta_{VB}/2\pi\rho_i$. 
 The small amplitudes below the transition range decreases the effectiveness of SH via the exponential suppression in Eq.~ (\ref{eq:DE}): moreover, the fact that $Q_{SH}$ is comparable to the overall cascade rate suggests that SH contributes to the steepening, with a considerable fraction of the turbulent cascade power heating the ions at these scales.
 %The steep spectral scaling of the fluctuations likely reduces the efficiency of the SH mechanism, which depends on large amplitude fluctuations. 
These results indicate that SH may be more effective at scales larger than the gyroradius \cite{Johnston2024}, contrasting studies that focus on SH at the thermal gyroscale \cite{Chandran2010,Vech2017,Martinovic2019,Cerri2021} (see end matter).%near the ion inertial scale $k d_i\sim1$ with $d_i=\rho_i/\sqrt{\beta_i}$and $\beta_i\approx 0.05.

%As $Q_{SH}$ approximately equals the turbulent cascade rate, and dominant over the ICW QL heating, the SH heating seems a potential source of the energy to drive the AI/C instability that likely generates the observed ICWs. Figure 3a) shows the normalized fraction of circularly polarized power, the contour showing the $\alpha <-4$ region as well as the frequency with maximum $dQ_{SH}$. Figure 3(b) shows the integrated left-hand polarized power against the total $Q_{SH}$ rate, with a ranked Spearman correlation $R_S=0.44$. We additionally plot in Figure 3(c), $Q_{SH}$ against $Q_{ICW}$, we find similar correlations.

\paragraph{Discussion}

 Stochastic mechanisms are important to a variety of plasma processes and environments \cite{Skilling1975,Drury1983,McChesney1987,McChesney1991,JohnsonCheng2001,Chen2001,Bourouaine2008,Hauff2009} including heating in the solar wind and corona. By extending the C+2010 SH formalism \cite{Chandran2010} through including both scale dependence \cite{Klein2016} and intermittency \cite{Mallet2019}, we demonstrate that heating via SH is largely consistent with observed turbulent cascade rates, as measured through local energy transfer \cite{MarschTu1997,Sorriso-Valvo2018}. Our focus on Alfvénic turbulence is motivated by widely realized observations of strong Alfvénicity of turbulent fluctuations in the inner-heliosphere \cite{Chaston2020,McManus2020,Shi2021,Bowen2025a}; however stochastic scattering via other coherent structures, e.g. pressure balance structures, magnetically dominated intermittent current sheets \cite{Bowen2018b}, and balanced turbulence \cite{Zhang2025} may also contribute to heating.  Our results indicate that stochastic magnetic moment breaking by intermittent structures in imbalanced turbulence is a viable heating mechanism in the sub-Alfv\'{e}nic solar wind, and provides important insight into coronal heating.
  
  PSP studies of turbulent heating at larger radial distances 
 %Contrasting previous studies of heating by PSP at larger radial distances, which 
 found net $\parallel$-ICW resonant heating on order of the turbulent cascade rate \cite{Bowen2022,Bowen2024b,Shankarappa2024}. Our present results indicate that, in contrast, in the sub-Alfv\'enic wind $\parallel$-ICWs are close to marginal stability, and, overall, cool the plasma.
Further work is needed to understand the transition, distribution, and difference between intervals where ICW heating is important \cite{Bowen2022,Bowen2024b,Shankarappa2024} versus those where it is negligible: the sub-Alfv\'{e}nic nature of the present interval may indicate a radial dependence on heating mechanisms, with $\parallel$-ICWs becoming more important in super-Alfv\'{e}nic regions. Further studies should investigate this transition, which may be the result of radial variations in the turbulence or $\beta$, and whether it supports signatures of preferential heating near the Alfvén surface \cite{Kasper2019}.
 
Importantly, we show empirically that the SH mechanism can only provide the relevant heating levels when intermittent effects are included; this point has been noted in theoretical work and simulations \cite{Mallet2019,Cerri2021}, though previous observational studies of SH \cite{Bourouaine2013,Vech2017,Martinovic2019,Martinovic2020} have not included intermittency. Our observations suggests that non-intermittent fluctuations may not be sufficiently large to drive SH.  Our results support the body of work that invokes dissipation by intermittent coherent-structures to explain observed inhomogenous heating \cite{Osman2011,Osman2012} that is associated with discontinuities \cite{Burlaga1991,Wu2013,Greco2012,Perrone2016,Qudsi2020,Sioulas2022a,Sioulas2022b,Perrone2023,Gonzalez2024}. It is well understood that intermittency significantly enhances cascade rates \cite{Karimabadi2013,Sorriso-Valvo2015}, such that turbulent dissipation must inherently be sensitive to intermittency. Our observed levels of SH, enabled by intermittency, provide an important connection of the dynamic process of magnetic moment breaking \cite{Chen2001,Bourouaine2008,Chandran2010,Hoppock2018} to the large body of work that empirically connects intermittency with collisionless heating, but where the specific dynamic mechanisms remain unknown. Indeed, our results in Fig \ref{fig:2} suggest that SH may be suppressed for the vast majority of non-intermittent fluctuations.%The specification of a specific dynamic process capable 

The interplay of linear and nonlinear processes is a contentious issue \cite{Coleman1968,Belcher1971,Howes2008,Schekochihin2009,Matthaeus2015,Oughton2020,Papini2021,Zhao2023} that is fundamental to understanding the turbulent mechanisms that shape astrophysical \cite{Verscharen2019,Howes2024} and laboratory \cite{Howes2018} plasmas. Our results clearly support the role of intermittency in turbulent dissipation. Recently, \cite{Johnston2024} proposed that SH and QL heating by oblique ICWs are related processes (with a similar $D_E$) that exist respectively within limits of balanced and imbalanced turbulence; additionally, they predict that heating in imbalanced turbulence should peak above the transition range, consistent with our observations, %. The increased levels of SH at scales larger than the thermal gyroscale (which is associated with the end of the range \cite{Bowen2020c,Bowen2024b,McIntyre2024}) is 
and in contrast with previous focus on SH at the thermal gyro-scale \cite{Bourouaine2013,Vech2017,Martinovic2019,Mallet2019,Cerri2021}. Further work is needed to explore similarities of cyclotron resonant and stochastic processes. Importantly, QL modeling \cite{Chandran2010b,Squire2022,Zhang2025,Johnston2024}, does not incorporate the effects of intermittency, which we show is likely necessary to reconcile observed turbulent cascade and heating rates. These insights are particularly interesting in the light of simulations suggesting that SH should primarily occur in balanced, low cross helicity turbulence, rather than the highly imbalanced, Alfv\'{e}nic, turbulence observed in this interval \cite{Zhang2025}, which could be subject to a helicity barrier \cite{Meyrand2020,Squire2022} type mechanism that may result in oblique cyclotron resonant heating. Understanding intermittent effects on QL cyclotron interactions \cite{Mallet2019} would help clarify this interplay. Additionally, while our turbulent cascade rates agree with the SH levels, empirical measurements of $D_E$ \cite{Arzamasskiy2019,Vasquez2020} and investigating $\alpha$-particle signatures of these mechanisms \cite{Isenberg1984,Hollweg2002,Chandran2013,Zhang2025} may help distinguish between (or, indeed, unify) QL processes and SH.

These results highlight several subsequent avenues for continued investigation. ICW dissipation has been shown to affect intermittency in sub-ion kinetic scale turbulence \cite{Bowen2024a} and it is unclear whether a dominant SH mechanism would have similar effects. In this same vein, studying the stochastic heating of electrons by kinetic scale current sheets \cite{Chasapis2015,Lotekar2022} may prove useful in understanding electron scale dissipation and intermittency \cite{Kiyani2009,Chhiber2022,Belardinelli2024}. Due to the relative infrequency of magnetic reconnection observed by PSP \cite{Eriksson2024}, which is likely a result of the Alfv\'{e}nic nature of inner-heliospheric turbulence \cite{Mallet2025}, we have mostly glossed over reconnection as a dissipation mechanism \cite{Mallet2017,Vech2018}. However, studying the interplay between SH and magnetic reconnection may be an important avenue in understanding the heating of the lower corona, where the relative importance of reconnection heating \cite{Parker1988,Drake2012,Raouafi2023a} and dissipation of Alfv\'{e}nic turbulence \cite{DePontieu2007,Cranmer2007,Ballegooijen2017,Uritsky2017} is still a matter of debate.

In conclusion, there is a growing consensus that intermittent heating is fundamental to turbulent dissipation and collisionless heating. These results provide a clear connection between inhomogeneous heating via intermittent dissipation and the dynamics associated with violation of the first adiabatic invariant, namely particle gyromotion, by turbulence. These results, which connect intermittent heating to Alfv\'{e}nic turbulence, may help unify understanding of collisionless heating across relatively disparate phenomenological descriptions of turbulence.

%The enhancement of SH heating through intermittency is consistent with These results potentially unify the intermittent heating by coherent structures with the dissipation of Alfv\'{e}nic turbulence via the SH mechanism.% that the intermittent While the original derivation of SH \cite{Chandran2010}

 %The simplified heating rate at the thermal gyroradius, $\rho_{th}$ from stochastic heating as developed in \cite{Chandran2010} \begin{equation}
%Q^{\rho_{th}}_{{SH}} =c_1\frac{\delta v_{\rho_{th}}^3}{\rho_{th}}e^{-c_2\frac{v_\perp}{\delta v_{\rho_{th}}}}\end{equation} which omits dependence on velocity space.

% which is in contrast to several previous studies \cite{Bowen2022,Bowen2024b,Shankarappa2024}) that found net ICW resonant heating on order the turbulent cascade rate. 

%s $Q_{SH}$ approximately equals the turbulent cascade rate, and dominant over the ICW QL heating, the SH heating seems a potential source of the energy to drive the AI/C instability
%\newpage

%\section{appendix}
\paragraph{Acknowledgements} TAB acknowledges  NASA grant 80NSSC24K0272 through the HSR program. TE acknowledges funding from The Chuck Lorre Family Foundation Big Bang Theory Graduate Fellowship and NASA grants NNN06AA01C and 80NSSC20K1285. OP acknowledges the project ``2022KL38BK -- The ULtimate fate of TuRbulence from space to laboratory plAsmas (ULTRA)'' (Master CUP B53D23004850006) by the Italian Ministry of University and Research, funded under the National Recovery and Resilience Plan (NRRP), Mission 4 -- Component C2 -- Investment 1.1, ``Fondo per il Programma Nazionale di Ricerca e Progetti di Rilevante Interesse Nazionale (PRIN 2022)'' (PE9) by the European Union – NextGenerationEU.
\providecommand{\noopsort}[1]{}\providecommand{\singleletter}[1]{#1}%

\section{end matter}
\paragraph{Survey of the Interval}
We consider the interval from PSP's eighth perihelion encounter from April 8th 2021 from 10:00:00-14:00:00 initially reported as the first encounter with the solar corona \cite{Kasper2021}. We use the merged PSP Search Coil and Magnetometer (SCaM) data \cite{Bowen2020b}, which was sampled at ~2kHz. Fig 4(a) shows a survey of the interval: the magnetic field in S/C coordinates; the magnitude of the field maintains a constant amplitude, which is consistent with large amplitude Alfv\'{e}nic turbulence \cite{Bowen2025a}. Fig 4(b) shows the SPANi velocity measurements in S/C coordinate system and frame. Fig 4(c) shows the total proton $\beta=v_{th}^2/V_A^2$ as well as the perpendicular and parallel contributions; uniformly $\beta \ll 1$, in agreement with the limits under which the SH diffusion in Eq.~(\ref{eq:DE}) is derived \cite{Chandran2010,Hoppock2018}. Fig 4(d) shows the Alfv\'{e}nic Mach number $U_{sw}/V_A$, where $U_{sw}$ is the solar wind speed computed in the heliocentric inertial frame. Fig 4(e) shows $\theta_{BV}$ the angle between the magnetic field and the solar wind velocity in the spacecraft frame for each measurement (the angle is taken between 0 and 90 degrees); the mean angle is $\langle\theta_{VB}\rangle=25^\circ$. Fig 4(f) shows the reduced magnetic helicity $\sigma_B$, \cite{HowesQuataert2010}, computed via a wavelet transform. The red band around 8-10 Hz  indicates left-hand circularly polarized ion-scale waves thought to be ICWs \cite{Bowen2020a,Bowen2020d}.
\begin{figure}[h]
    \centering
    \includegraphics[width=3.25in]{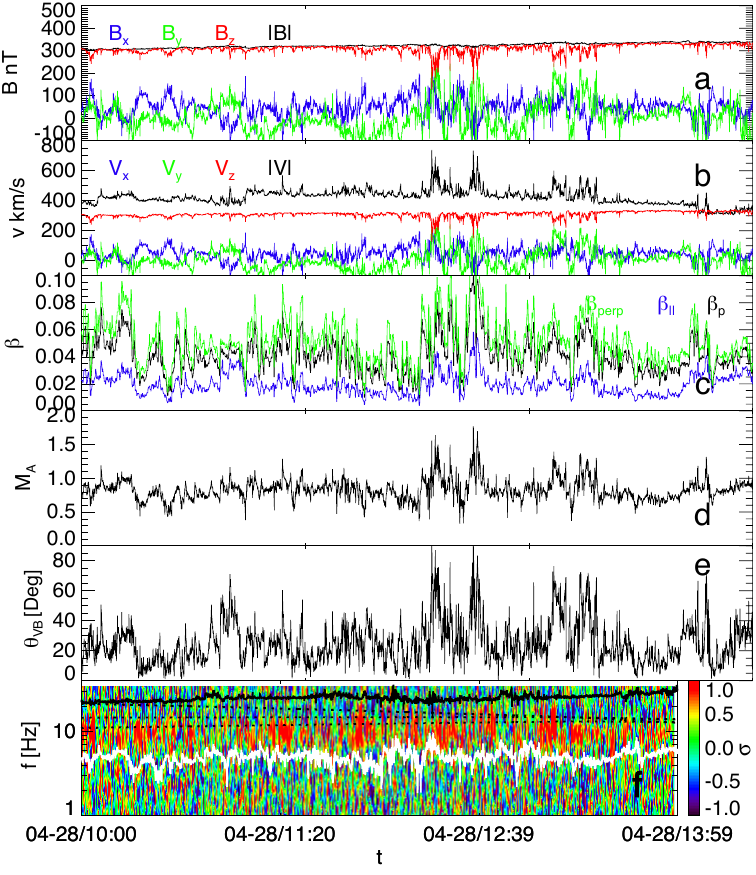}
    \caption{a) PSP FIELDS magnetic field in spacecraft coordinates on 2021-04-28 from 10:00:00-14:00:00. b) PSP SPANi measurements of flow velocity in S/C coordinates, c) Plasma $\beta$, total value  as well as perpendicular and parallel decomposition are shown. d) Alfv\'{e}nic Mach number. e) Angle between magnetic field direction and solar wind flow direction $\theta_{VB}$. f) Reduced magnetic helicity as a function of frequency and time; dashed lines show spacecraft reaction wheels, white and black lines show $kd_i$ and $k\rho_i$ respectively.}
    
    \label{fig:enter-label}
\end{figure}

\paragraph{Wavelets}
We use a 6th-order Morlet wavelet transform \cite{TorrenceCompo1998,DudokdeWit2013} to identify intervals with circular polarization via $\sigma_B.$ The full 3D vector fluxgate magnetometer measurement (sampled at 292.969 samples /sec) is used to compute the wavelet transform in field aligned coordinates, $\tilde{\mathbf{B}}(t,f)=(\tilde B_{\perp1},\tilde B_{\perp2},\tilde B_{\parallel})$. We follow methods developed in \cite{Bowen2022,Bowen2024b} to measure QL $\parallel$-ICW heating rates based on in situ observations of VDFs and wave populations. The reduced magnetic helicity \citep{HowesQuataert2010} is computed from the wavelet transform as
\begin{align}
        \sigma_B(f,t)=-2\text{Im}(\tilde{B}_{\perp1}\tilde{B}_{\perp2}^*)/(\tilde{B}_{\perp1}^2+\tilde{B}_{\perp2}^2).
\end{align} is used to find highly circularly polarized fluctuations, which are likely ion-scale waves propagating parallel the mean magnetic field and away from the sun. The expression for $Q_{ICW}$ is found in Eq A1 \cite{Bowen2024b}, the reader is guided to \cite{Bowen2022,Bowen2024b} for reproducing $Q_{ICW}$ here.

%\begin{align}\label{eq:H}
%\begin{split}
 %Q_{\rm ICW}=
% \frac{\pi m_i \Omega_i^2}{4}\int d^3\mathbf{v}\Bigg\{v^2 \int_0^\infty  dk_\parallel\frac{1}{v_\perp}&\\ \times\hat{G}_k\Bigg[ v_\perp \delta(\omega_k -k_\parallel v_\parallel-\Omega_i)\frac{\omega_k^2}{k_\parallel^2c^2} &I(k_\parallel)\hat{G}_k f(v_\perp,v_\parallel)\Bigg]\Bigg\}, 
%\end{split}
%\\
 %   \hat{G}_k=  (1- \frac{k_\parallel v_\parallel}{\omega_k})\frac{\partial}{\partial{v_\perp}} +\frac{k_\parallel v_\perp}{\omega_k}\frac{\partial}{\partial{v_\parallel}}
%\end{align} 
The spectrum of ICW is determined by filtering highly circularly polarized coefficients with $\sigma_B >0.7.$
%\begin{align}
%I(k_\parallel)=\frac{\tilde{B}^2_{LH\sigma}}{B_0^2}\frac{df}{dk_\parallel}.\end{align} 
%a scale dependent fluctuation amplitude of the magnetic field $\\delta{B}$ from the PSP Search Coil and Magnetometer (SCaM) data \cite{Bowen2020b} using a 6th-order Morlet wavelet transform \cite{TorrenceCompo1998,DudokdeWit2013}. 
%We calculate a scale dependent fluctuation amplitude of the magnetic field $\\delta{B}$ from the PSP Search Coil and Magnetometer (SCaM) data \cite{Bowen2020b} using a 6th-order Morlet wavelet transform \cite{TorrenceCompo1998,DudokdeWit2013}. 
We compute drifting biMaxwellian fits to the SPANi VDFs, assuming a core (c) and drifting beam (b) biMaxwellian populations 
\begin{align} 
f(v_\perp,v_\parallel)=\sum_{j=b,c}\frac{n_j \pi^{-3/2}}{w_{j,\perp}^2 w_{j,\parallel}}\text{exp}\left[-\frac{v_\perp^2}{w_{j,\perp}^2} -\frac{(v_\parallel-v_0)^2}{w_{j,\parallel}^2}\right]\end{align}
 using nonlinear least square fitting methods \cite{Verniero2020,Klein2021,Bowen2022}. We explored radial basis function interpolation to account for non Maxwellian effects, but found no significant changes\cite{Bowen2024b}.

\paragraph{5-Point Increments}
Two point increments cannot resolve the steep scaling of the transition range \cite{Cho2009,Frisch1995} and thus we use 5-point increments \cite{Cho2019} to compute fluctuation amplitudes as \begin{align}
\delta B_i(t,\tau) =\Delta_{5pt}B= &\\\frac{1}{\sqrt{35}}[B_i(t-2\tau )-&4B_i(t-\tau)+6B_i(t)\\&-4B_i(t+\tau)+B_i(x+2\tau)].\end{align}  where $i$ refers to the vector components. The KAW correction converts each increment $\delta B_i$ to a  $\delta v_i$.

Only two components of the SCM are available such that the total instantaneous amplitude is computed as $\delta{u}(f,t) =\sqrt{3/2( \delta u_v^2  +\delta u_w^2) }$ where subscripts $v$ and $w$ denote the functional SCM axes.
The rms $D^{rms}_E$ is computed
from the time averaged rms $\delta u^{rms}(f)=\sqrt{\langle |\delta u(f)|^2\rangle}$. Alternatively $D_E(t,f)$ is computed through $|\delta u(f,t)|$ in Eq.~(\ref{eq:DE}). Evaluating both $D^{rms}_E$ and $\bar{D}_E$ in a scale dependent manner additionally requires determining $v_\perp(f)$. A primary observational complication in this work is associating $\delta u(f,t)$ with the appropriate $v_\perp(f)$.

Our results show that heating mostly occurs near the ion-transition range, which is well resolved by the fluxgate magnetometer (MAG), for which 3D vector measurements are possible. To verify the results from the the 2-axis SCaM data, we additionally repeated the analysis with the MAG data. No significant changes were found in the total heating rate, as can be seen in Figure 5(a).

\paragraph{Modified Taylor Hypothesis}
The Doppler shift equation is generally given by \begin{align}
    2\pi f=\omega(\mathbf{k})+\mathbf{k\cdot V_{sw}}.
\end{align} The Taylor hypothesis is obtained simply by dropping the $\omega(\mathbf{k})$ dependence. The modified Taylor hypothesis in Eq.~(\ref{eq:modtay}) should be appropriate for the large-amplitude Alfv\'{e}nic fluctuations, propagating with $\omega=k_\parallel V_A$\cite{Klein2015,Bourouaine2019,Chhiber2019}. The hypothesis is valid when the amplitude of the subdominant Elsasser mode is less than the spacecraft's perpendicular speed in the heliocentric inertial frame $|\delta\mathbf{z}^-|\ll V_{sc\perp}$ and additionally $|\delta\mathbf{z}|\ll|\delta\mathbf{z}^+|$
\cite{Klein2015}, which are generally satisfied in the stream under study.

The condition $k_\perp\rho=1$ 
corresponds to $k_\perp= \Omega_i/v_\perp$. Following the modified Taylor Hypothesis, with the assumption of critical balance
$k_\parallel V_A=\delta u \Omega_i/v_\perp,$ we find

\begin{align} \label{eq:modtay}
 %v_\perp=\frac{\Omega_i}{2\pi f }\left(\frac{\delta u}{v_\perp} +\frac{\delta u}{v_\perp}\frac{V_{sw}}{{V_A}} cos\theta_{VB}+\frac{1}{v_\perp} V_{sw} sin\theta_{VB}\right).
%2\pi f_{str}= k_\perp V_{sw} sin\theta_{VB}\\
v_\perp=\frac{\Omega_i}{2\pi f }\left[{\delta u}\left(1+ \frac{V_{sw}}{{V_A}} cos\theta_{VB}\right)+ V_{sw} sin\theta_{VB}\right].
%2\pi f_{str}= k_\perp V_{sw} sin\theta_{VB}\\
\end{align}

where we use $\delta u^{rms}(f)$ to convert from $f$ to $v_\perp$.

We also consider a purely two dimensional (2D) perpendicular spectra with $k_\parallel=0$
$$v^{2D}_\perp=\frac{\Omega_i}{2\pi f } V_{sw} sin\theta_{VB};$$
 an isotropic spectra $k_\perp=k_\parallel$, which yields
$$v^{iso}_\perp=\frac{\Omega_i}{2\pi f}\left( V_A+V_{sw}cos\theta_{VB}+V_{sw} sin\theta_{VB}\right);$$
and the standard Taylor Hypothesis (TH)
$$v^{TH}_\perp=\frac{\Omega_i}{2\pi f}V_{sw}.$$

\begin{figure}
    \centering
    \includegraphics[width=\columnwidth]{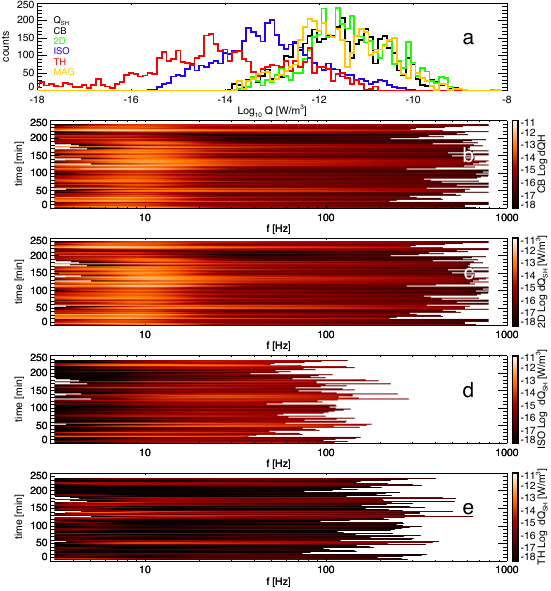}
    \caption{a) PDF of heating rates for various assumptions of spectral anisotropy: CB (black), 2D (green), isotropic (blue), standard Taylor (red). b-e) respective estimates of $dQ_{SH}$ for assumptions (CB, 2D, ISO,TH) of spectral anisotropy.}
    \label{fig:a2}
\end{figure}

Figure \ref{fig:a2} shows results using various definitions of $v_\perp$ based on different assumptions of the spectra. Assuming an entirely perpendicular 2D spectra yields similar heating rates as a CB spectra; the use of isotropic or a standard TH give much lower heating rates. Figure \ref{fig:a2}(b-d) show the corresponding $dQ_{SH}$ calculations for the CB, 2D, isotropic, and standard TH analysis. The CB is the same as in Figure \ref{fig:3}, but replotted on the same scale for comparison with the other spectra. The clear similarities in $dQ_{SH}$ between the CB and 2D data, and obvious differences in comparison with isotropic spectra and standard TH approaches, suggest that our analysis should hold as long as the turbulent spectra are strongly perpendicular.
\paragraph{LET}

To estimate the total energy cascade rates $\epsilon^T$ we use the local-energy transfer \cite{MarschTu1997,Sorriso-Valvo2015,Sorriso-Valvo2018}. Standard Politano-Pouquet \cite{PolitanoPouquet1998} estimates of the magnetohydrodynamic cascade rate require mixed 3rd order moment of increments of the Elsasser variables $\Delta\mathbf{z}^{\pm}=\mathbf\Delta{v}\pm\Delta\mathbf{B}/\sqrt{\mu_0\rho_0}$ given as 
\begin{equation}
    \langle\epsilon^\pm\rangle=-\frac{3}{4}\langle|\Delta\mathbf{z}(t,\tau)^\pm|\Delta z(t,\tau)^{\mp}\rangle\frac{1}{V_{sw}\tau}.
\end{equation}

The unaveraged $\epsilon^T=(\epsilon^++\epsilon^-)/2$ can be interpreted as an estimate of the local cascade rate associated with individual increments \cite{Sorriso-Valvo2018}. The distributions in Figure \ref{fig:1}  show distributions of the local cascade rate at three lag values (3.5, 49, and 168 seconds) corresponding to the inertial range over the entire 4 hour interval. The distributions are roughly consistent with the dissipation associated with intermittent stocahstic heating. Intermittency has been shown to directly contribute to non-homogeneity of the turbulent cascade rate \cite{Sorriso-Valvo2015}, which is similar to our present results that highlight the importance of intermittency in stochastic heating.

\end{document}